# Priority Based Energy-Efficient Data Forwarding Algorithm in Wireless Sensor Networks

Deepali Virmani  
Assistant Professor, Department of CSE  
Bhagwan Parshuram Institute of Technology  
Sector-17, Rohini, Delhi-110089, India  
deepalivirmani@gmail.com

Dhruv Talwar  
Department of IT  
Bhagwan Parshuram Institute of Technology  
Sector-17, Rohini, Delhi-110089, India  
dhruv.talwar31@gmail.com

Arun Dhingra  
Department of IT  
Bhagwan Parshuram Institute of Technology  
Sector-17, Rohini, Delhi-110089, India  
dhingra.arun.18@gmail.com

Tushar Bahl  
Department of IT  
Bhagwan parshuram Institute of Technology  
Sector-17 , Rohini, Delhi-110089,India.  
tush.bahl@yahoo.in

*Abstract*— A Wireless Sensor Network (WSN) can be described as a collection of untethered sensor nodes. An important application of WSNs is in the field of real-time communication. Real-time communication is a critical service which requires a qualitative routing protocol for energy-efficient network communication. The judicious use of energy of the network nodes is essential and important for sustainability and longevity of a WSN. This paper proposes an algorithm namely 'Priority-Energy Based Data Forwarding Algorithm(PEDF)' which empowers the node to choose the most suitable packet forwarding path, based on the priority of the packet and the current energy status of the forwarding node. The algorithm hence dynamically adapts to the prevailing energy-scenario of the network and takes routing decisions accordingly, based on packet priority. Minimizing delay, minimizing energy utilization, maximizing throughput and maximizing network lifetime are the key elements of the proposed algorithm.

*Keywords*— Wireless sensor networks; priority; energy, real time.

## I. INTRODUCTION

A sensor is a device that measures a physical quantity like temperature, pressure, pollutants etc and converts it into a signal which can be read by a dedicated instrument. Sensor nodes are small, low-cost, low-power, multifunctional devices which come together to form a sensor network [2].

A Wireless Sensor Network (WSN) is composed of a large number of sensor nodes (not connected by wires), which coordinate together to perform some specific actions [1].

Supporting real-time communication in WSNs is a challenging issue. Sensor nodes carry limited, generally irreplaceable, power sources due to their inexpensive nature and ad-hoc method of deployment. So, one of the most important constraints on sensor networks is to minimize energy consumption[2].WSN applications (e.g., border surveillance) must operate for months without wired power supplies. Therefore, WSNs must meet the delay requirements of packets at minimum energy cost. For instance, authorities need to be notified sooner about high-speed motor vehicles than slow-moving pedestrians. To support such applications, a real-time communication protocol must adapt its behavior based on packet deadlines.

For transmitting the data packets from one source node (S) to another destination node (D), many transmission paths may be available with varying delay parameters involved. So as to send the data packet in least time period, the routing protocols often uses the minimum time utilization path between S and D node. This path is known as the Best Path. While choosing the best path, the routing protocols usually neglect the current power levels of the nodes in the path. This ignorance of power availability of nodes leads to repetitive and continuous use of the best path even for sending unimportant and non-critical packets. As a result, after some time this path will get out of power and will be of no further use to the WSN. In such a case, packets especially critical ones cannot use the best path and will have to take other paths, incurring large unacceptable time delays. For example, an alert for fire in the house requires the use of best path more than an alert on water leakage from a tap. This may possibly lead to an early collapse of the network and its purposes.

In order to deal with such problems relating to early node and network energy depletion and accessibility of best path by different packets of varying criticality, we propose an efficient algorithm for forwarding data packets considering their priority and the power availability of the nodes in the best path. This ensures that there is no over exploitation or energy exhaustion of the best path and it is always available for transmission of critical data packets, timely and speedily.

## II. LITERATURE REVIEW

In [1], the author talks about ad hoc deployment of sensor nodes. There are large numbers of untethered and unattended sensor nodes, which have to operate for years. The wireless sensor nodes are furnished with limited power source. Thus the optimal use of power/energy of nodes is of prime importance.

Also, the sensor nodes may be equipped with effective power scavenging methods such as solar cells, because the sensors may be left unattended for months and even years [2].

Another type of network is the WSAN (Wireless Sensor Actor Networks) as described in [3]. WSANs comprise of sensors and actors, where sensors gather information about the physical world, while actors take decisions and then perform appropriate actions upon the environment. It allows a user to effectively sense and act from a distance.

SPEED[4] deals with end to end real time communication by preserving the same speed across the network using non-deterministic geographic forwarding which can balance traffic and reduce congestion in a bandwidth-limited network.

In [5], the author proposes a protocol MMSPEED, which focuses on a delivery mechanism for improving the QoS (Quality of Service) namely timeliness and reliability, in wireless sensor networks MMSPEED provides multiple delivery speed options that are guaranteed network wide. MMspeed extends the SPEED to support different delivery velocities and levels of reliability.

Both, [4] and [5] focus on providing appropriate delivery speeds to the data packets. Although, this increases the number of packets meeting their deadlines but no consideration is given to the power levels of the nodes present in the WSN.

RPAR[6] supports energy efficient real time communication in wireless networks by adapting the power and routing decisions based on packet deadlines. It is based on a tradeoff between transmission power and delay. When the deadlines are strict it trades the energy and capacity for decreasing delay by increasing the transmission power. Conversely, when the deadlines are loose, it lowers the power to reduce energy consumption.

## III. PROBLEM DOMAIN

SPEED and MMSPEED routing protocols [4][5], do not pay any attention to the current energy levels and remaining battery life of the nodes present in the network. This may result in early and untimely power exhaustion of the network either partially or completely, thus drastically reducing the serving life of the network. This is not favourable.

On the other hand, to ensure timely packet delivery, RPAR [6], takes into account the energy levels of nodes for selecting the best forwarding node. But, in the process it tends to over-use a particular node or the whole forwarding path. This is because it uses the same best path to reach to the destination node every time, irrespective of the priority or importance of the data packet. As a result, the best path nodes get exploited even by the non-critical data packets. This may lead to quick energy level depletion of the best path nodes, eventually leading to their total energy exhaustion, making them futile for further use.

To further aggravate this problem, if in such a situation a packet of high urgency (like to inform the authorities about theft or fire) is to be transmitted from one node to another, then the best path is not available to it (as its exhausted).So, the critical packet will have to choose the second best path, incurring unacceptable time delay in transmission.

Such exploitation of best path by not-so-important packets and as a result hindering the use of the best path by urgent packets is not favourable. This gravely affects the longevity of an efficient wireless sensor network.

The above scenario is explained below in figure 1 which gives a snapshot of a network's portion with 9 nodes. Here 'S' and 'D' are the source and destination nodes of the data packet respectively. The packet is first transmitted through the best available forwarding path from S to node 2, via node 1. Now, we consider that the best path to reach D from node 2 is via node 3 (shown by solid arrows), ensuring speedy and timely delivery. But if node 2 repetitively forwards packets via this best path even for non-critical packets it will lead to early energy depletion of the path, eventually making it unusable. As a result, all packets especially the urgent ones will have to take the second best path to D via node 4 (shown by dashed arrows) which may cause unnecessary and undesirable delay in delivery of packets

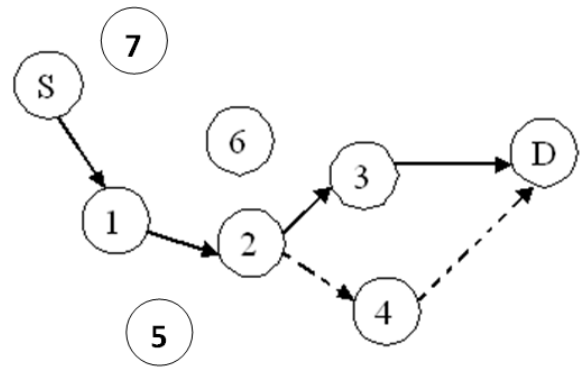

Figure 1. Transmission path of nodes

So we conclude that the best path should be available at all times to the urgent packets and should be allowed for use by unimportant packets only depending on the path's power status.

## IV. PROPOSED ALGORITHM

Considering the problem domain, we need to conserve the energy of the best path nodes to ensure their availability at all times especially to urgent data packets.

This can be done successfully by prioritizing the importance of the data packets, depending upon the urgency of the data to reach the destination node from the source node.

We, therefore suggest the following techniques and then an algorithm as the solution for this priority problem.

### A. Priority Assignment

All data packets need to be assigned a priority by the source node, depending on how important is it for the packet to reach the destination node. The priority so decided by the source node should be stored in a field, named 'Priority', of the data packet's header and should remain a part of the header, untampered, till it reaches the destination node.

The source node uses the 'Priority-Assignment Table', shown in table 1, to assign one of the four priorities to the packet depending on their urgency levels.

TABLE 1. PRIORITY-ASSIGNMENT TABLE

| Priority Level | Importance |
|---|---|
| Priority 1 | Urgent |
| Priority 2 | Highly Important |
| Priority 3 | Moderately Important |
| Priority 4 | Less Important |

### B. Forwarding Choice Determination

As all data packets always possess a header field of 'Priority', all the path nodes can read the packet's priority field and depending on that take appropriate forwarding decisions. Along with the priority, the node must also know the current energy levels of the forwarding node, so as to decide whether the existing power status allows it to forward the given priority data packet or not.

This decision is made by the nodes by using the below stated 'Priority-Energy Table' in table 2.

TABLE 2. PRIORITY-ENERGY TABLE

| Case | Percentage Energy Level of the Node | Allowance for Packets Having |
|---|---|---|
| Case I | 0-25% | Priority 1 |
| Case II | 25-50% | Priority 1 and 2 |
| Case III | 50-75% | Priority 1, 2 and 3 |
| Case IV | 75-100% | Priority 1, 2, 3 and 4 |

For example, in fig. 1, let us consider that node 2 has to forward a data packet of priority 4 (set by S) and it knows that the current energy level of node 3 is 27%.

Then in such a case, node 2, after consulting table II (Case II), will forward the packet to the second best path i.e. to node 4. Also, it would have forwarded a packet of priority 3 to node 4 only, under the same case. But, node 2 would have forwarded priority 1 and 2 data packets via the best path i.e. via node 3.

This is because node 2 knows that node 3 has fairy good amount of power to forward data packets of priority 1 and 2, as they are urgent. On the other hand, node 3 is currently not that power-rich (27% energy) to forward less critical i.e. priority 3 and priority 4 data packets, which if forwarded will further deplete its power.

Similarly, in case, node 3 has more than 75% energy left, then node 2 will forward all priority data packets via the best path i.e. via node 3, as it currently has enough power to forward all data packet (Case IV).

As more and more packets will be forwarded via node 3, its power will get depleted, giving rise to different cases as mentioned in table 2 and accordingly, forwarding decisions will be taken.

As a result of this, the best path nodes are saved from getting power depleted by non-urgent data packets and so the best path always remains available for critical packets' transmission.

### C. Power-Status Reporting

An important aspect of this technique is that how will node 2 know about the current energy level of node 3(refer fig 1), as it has to take a packet forwarding decision based on node 3's energy status. Its solution is power–status reporting by node 3 to its neighbor nodes (like node 2). Node 3 reports its power, whenever it reaches an energy level, just above, below or the exact value of the critical energy values of 75%, 50% and 25% of the total node energy.

Initially, all nodes have 100% power. So, according to this reporting scheme, node 3 should inform node 2 about its power-status as soon as its power level depletes to 75% or less. Similarly, node 3 should report its power-status to node 2 at other two critical values of 50% and 25%.

As a result of such reporting, a node always knows which of its neighbor nodes have what energy levels and accordingly, it makes the choice of the best forwarding node (based on table II) for a particular priority data packet.

### D. Power Replenishing

The above proposed technique of forwarding packets, based on packet priority and node energy, becomes even more significant with the presence of energy or power replenishing entities like solar cells in the node architecture.

As a result, if a node depletes its power level below 25%, its workload gets automatically decreased (as it doesn't have to forward packets of priority 2, 3 and 4), giving it ample time to replenish its power using its solar cells. As its power level increases above critical values, it informs its neighbor nodes under the power-status reporting scheme and gets ready for more workload.

Based on the above explained schemes and techniques, we propose the following 'Priority-Energy Based Data Forwarding Algorithm' in figure 2.

---

**Algorithm : PRIORITY-ENERGY BASED DATA FORWARDING ALGORITHM**

START

  Initialize p : Priority of data packet from packet header ;

  Switch(p)

    Case 1 : Find ANY suitable neighbor node for packet
         forwarding ;
         Exit switch ;
    Case 2 : Find ONLY those suitable neighbor node
         for packet forwarding having energy
         level greater than 25% ;
         Exit switch ;
    Case 3 : Find ONLY those suitable neighbor node
         for packet forwarding having energy
         level greater than 50% ;
         Exit switch ;
    Case 4 : Find ONLY those suitable neighbor node
         for packet forwarding having energy
         level greater than 75% ;
         Exit switch ;
    default : Request retransmission of packet from
         previous node ;

  Exit Switch ;

END

---

FIGURE 2. PRIORITY-ENERGY BASED DATA FORWARDING ALGORITHM

## V. CONCLUSION

Energy conservation is an important aspect for real-time routing in WSNs. Not many protocols focus on judicious use of network power. So in this paper, we proposed and established a relationship between packet priority and node energy in the form of the 'Priority-Energy Based Data Forwarding Algorithm', which decides on the best energy-efficient forwarding choice. Also, the proposed PEDF emphasis on reducing the workload on some specific nodes, depending on their power levels, and giving them time for replenishing their energy. In the process, no particular node gets over-exploited and the herculean task of smooth and efficient network functioning and management is more or less equally distributed among all the nodes. Therefore, it ensures a proficient real-time communication in WSN, with minimizing the delay and maximizing the energy utilization which further maximizes the network lifetime. Further we can simulate and compare the proposed work with the existing scenarios and can prove its validity.